# Public Auditing and Data Dynamics for Storage Security in Cloud Computing

Hemalata A. Gosavi [#1], Prof. Manish R. Umale [*2]

*# ME Student, Department of Computer Engineering.*

*Lokmanya Tilak College of Engineering*

*Vikas Nagar, Sector 4, Koperkhairane, Navi Mumbai.India.*

*Abstract*— Cloud computing has been envisioned as the next-generation architecture of IT Enterprise. Using Cloud Storage, users can remotely store their data and enjoy the on-demand high quality applications and services from a shared pool of configurable computing resources, without the burden local copy data storage and maintenance [1]. It moves the application of software data stored to the centralized large data centres, where the management of the data stored services may not be completely trusted [1]. There are many new security challenges and the problems taken into account for ensuring the integrity of data storage in Cloud Computing [1]. In particular, we consider the task of allowing a third party auditor (TPA) to perform verifies the integrity of the dynamic data stored in the cloud [1].

## I. INTRODUCTION

Cloud Computing is the long dreamed vision of computing as a utility, Cloud Computing is used to store the large amount of data in the remote space or remote location. Using the cloud space area; we can remotely access the data and enjoy the on-demand high quality application and services from the remote location. In Cloud Computing, we can share remote data with our proposed system. As we store local data in the remote location, our local servers are free from the work burden. By data outsourcing, users can be relieved from the burden of local data storage and maintenance. As we store local data in the remote location, our local servers are free from the work burden. To securely introduce an effective TPA in the auditing process TPA perform audits for multiple users simultaneously and efficiently. Extends security and performance analysis show the proposed system are provably secured and highly efficient data stored on cloud. [1, 4]

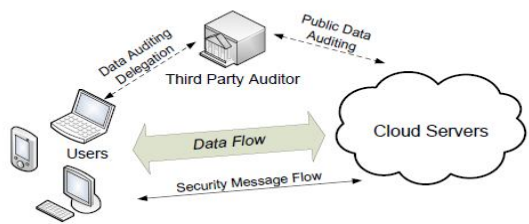

Fig. 1 the Architecture of Cloud Data Storage Service

## II. PROBLEM STATEMENT

We consider cloud computing a cloud data storage service involving three aspects are the cloud user (client) who has large amount of data files to be stored in the cloud; the cloud server (CS), which is managed by the cloud service provider (CSP) to provide data storage service and has large storage space and computation resources; the third party auditor (TPA), who has expertise and capabilities that cloud users do not have and is trusted to assess the cloud storage service reliability on behalf of the user upon request. Users depend on the CS for cloud data storage and maintenance. They might likewise dynamically interface with the CS to get to and update their put away data for different provision purposes. As clients no more have their data provincially, it is of discriminating significance for clients to ensure that their data are being correctly stored and maintained. To save the calculation asset and the online burden potentially brought by the periodic storage rightness confirmation, cloud clients may fall back on TPA for ensuring the storage integrity of their outsourced data, while planning to keep their data private from TPA.

## III. SYSTEM DESCRIPTION

### A. System Models

We assume that the system is composed of the following parties: the Data Owner, Cloud Servers, and a Third Party Auditor. To access data files shared by the client, data owner, Data Consumers, or users for quickness, download data files of their interest from Cloud Servers and then decrypt. The data owner and users will not to be constantly online. They come online simply on the need premise. For simply, we expect that the main access benefit for clients is data file reading. Extends our proposed scheme to support data file writing is trivial by asking the data writer to sign the new data file on each update as does. we will also call data files by *files* for quickness. Cloud Servers are always online and operated by the Cloud Service Provider (CSP). They are expected to have abundant storage capacity and computational power. The Third Party Auditor (TPA) is also an online party which is used for auditing time every file access event. In additional, we also assume that the data owner can not only store data files but also run his own code on Cloud Servers to manage his data files.

*B. Security Models*

In this work, Cloud Servers will follow our proposed system in general, but try to find out as much secret data as possible based on their inputs. More specially, we expect Cloud Servers are more interested in file contents and user access privilege information than other secret data. Cloud Servers might collide with a small number of malicious users for the purpose of harvesting files contents when it is high beneficial. Communication between the data owner/users and Cloud Servers are assumed to be secured under existing security system. Users would try to access files either within or outside the scope of their access privileges. To achieve this goal, authorized users may work on own data independently to store data insert data, delete data, modify data on cloud.

*C. Design Goals*

Our main design goal is to help the client achieve fine-grained access control on files stored by Cloud Servers. Specially, we need to enable the client (data owner) to enforce a unique access structure on each user, which accurately design at the set of data files that the user is allowed to access. We also need to prevent Cloud Servers from being able to learn both the data file contents and user access benefit data. In additional, the proposed scheme should be able to achieve security goals like user auditing and support basic operations such as block insertion, block modification, block deletion, block modification, block verification and taking client log history also the user grant as a general one-to-many communication system would require. All these design objective should be achieved efficiently in the manner that the system is scalable.

IV. MODULE DESCRIPTION

*1. Third Party Auditor (TPA):*

TPA performs reviews for multiple clients simultaneous and efficient. Extends security and performance analysis demonstrate the proposed framework are provably secured and very effective [1, 4]. To securely introduce a powerful third party auditor (TPA), the accompanying two essential prerequisites must be

1) TPA should be able to effectively audit the cloud information storage without requesting the neighbourhood duplicate of data, and present no extra on-line burden to the cloud client;

2) The third party auditing procedure should to acquire no new vulnerabilities towards client data privacy. In this framework, we use and particularly consolidate the general public key based homomorphic authenticator with random masking to achieve the privacy-preserving public cloud data auditing system, which meets all above prerequisites. To help efficient handling of multiple auditing tasks, we further explore the method of bilinear aggregate signature to extend our main result into a multi-user setting, where TPA can perform various auditing tasks simultaneous. Extended security and performance analysis indicates the proposed plans are provably secure and highly efficient.

*2. Privacy-Preserving Public Auditing Module:*

Homomorphic authenticators are reprehensible verification metadata produced from singular data blocks, which might be securely aggregated in such an approach to guarantee an auditor that a linear combination of data blocks is effectively registered by verifying only the aggregated authenticator. Overview to attain privacy-preserving public auditing, we propose to particularly coordinate the homomorphic authenticator with random mask technique. In our convention, the linear combination of sampled blocks in the server's response is masked with irregularity produced by a pseudo random function (PRF).

The proposed scheme is as takes after:
1. Setup Phase
2. Audit Phase

*3. Batch Auditing Module:*

With the foundation of privacy-preserving public auditing in Cloud Computing, TPA might simultaneously handle various auditing delegations upon distinctive clients' appeals. The singular auditing of these tasks for TPA could be tedious and extremely wasteful. Batch auditing not just permits TPA to perform the different auditing tasks at the same time, additionally significantly reduces the calculation cost on the TPA side.

*4. Data Dynamics Module:*

Supporting data dynamics for privacy-preserving public risk auditing is additionally of paramount importance. Presently we indicate how our fundamental scheme could be adjusted to expand upon the current work to help data dynamics, including block level operations of block modification, block deletion and block insertion. We can receive this system in our outline to attain privacy-preserving public risk auditing with support of data dynamics.

Data dynamics implies after clients store their data at the remote server, they can dynamically upgrade their data at later times. At the block level operation, the principle operations are insertion, modification, deletion, verification, and taking log history.

**Block Insertion**: The block insertion operation server can insert anything on the existing client's file or new introduce client file.

**Block Deletion**: The bloc deletion operation server can delete anything on the client's file.

**Block Modification**: The block modification operation server can modify anything on the client's file.

**Block Verification:** The block verification operation TPA gives acknowledgement block is modified or not modified.

**Log History**: The client performs basic block operation on client's file taking log details.

V. MODELING THE SYSTEM

A public auditing consists of four techniques (KeyGen, SigGen, GenProof, and VerifyProof).

**KeyGen:** key generation technique that is run by the client to setup the scheme. Key generation work as the client initializes the public and secret parameters of the system. Client uploads the data files in the encrypted format and decrypts its own data only.

**SigGen:** Sign generation utilized by the client to produce verification metadata, which may comprise of digital signatures or other data utilized for auditing. The client stored data files on the cloud server and verify metadata, and deletes local copy on client.

**GenProof:** Generation proof run by the cloud server to generate a proof of data storage correctness. GenProof after executing gives the acknowledgement its verification metadata input.

**VerifyProof:** Verify proof run by the TPA to audit the proof from the cloud server. VerifyProof also gives the acknowledgement via TPA verifies the metadata.

VI. EXPERIMENTAL ANALYSIS

We now show report on experimental result of our experiments. In our experiments as shown in Fig.2 clients upload data files on cloud server and delete the local copy. Cloud server verifies the inputs of metadata and TPA also verifies and give the acknowledgement to sever data is modified or not.

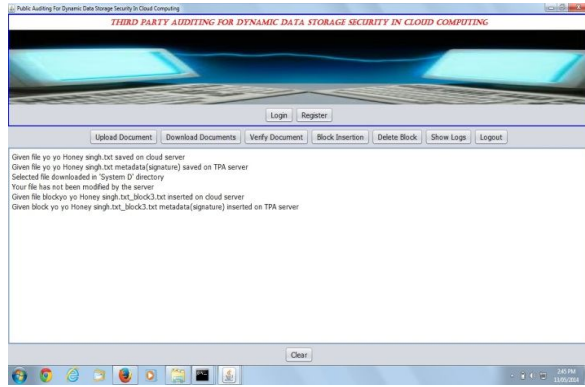

Fig.2 shows Block insertion Operation

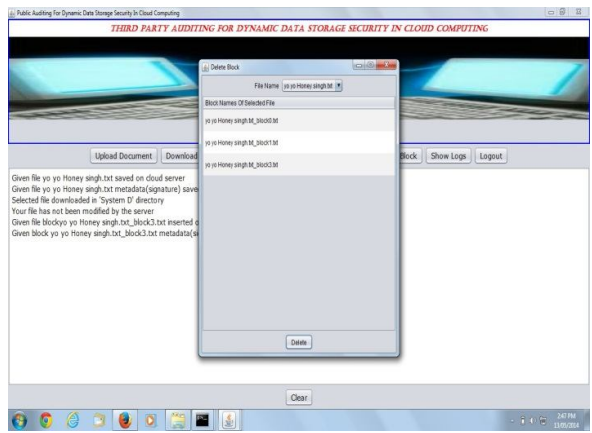

Fig.3 shows Block metadata on client.

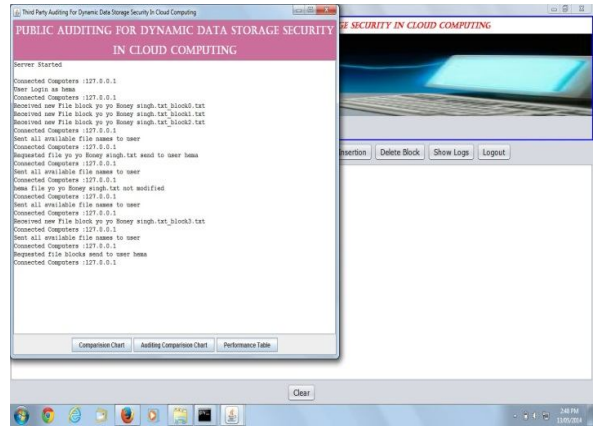

Fig.4 shows Block metadata on TPA.

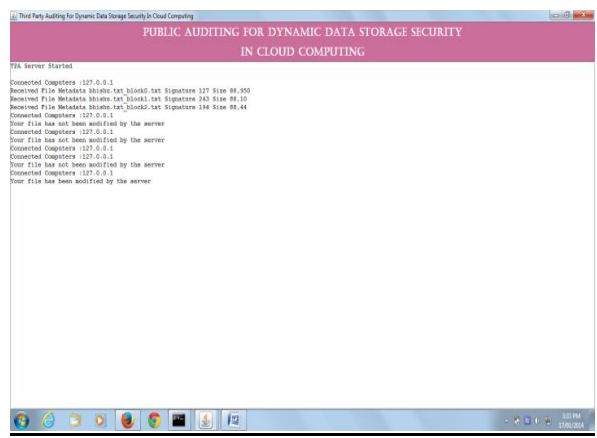

Fig.5 shows Client log history with operation.

Fig.6 TPA verifies and give acknowledgement data not modified.

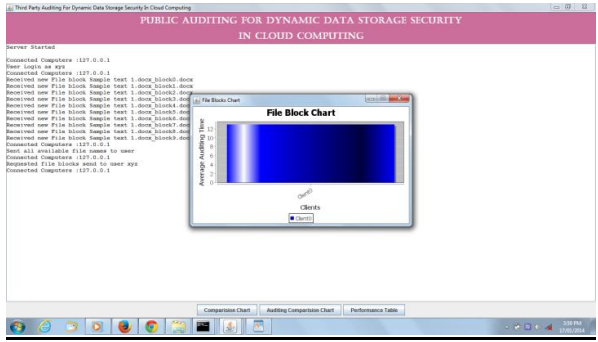

Fig.7 shows metadata into the block and client.

As shown in Fig.3 the data files converts in to number of Data blocks and client details. Similarly Fig.4 shows the block auditing per block is required.

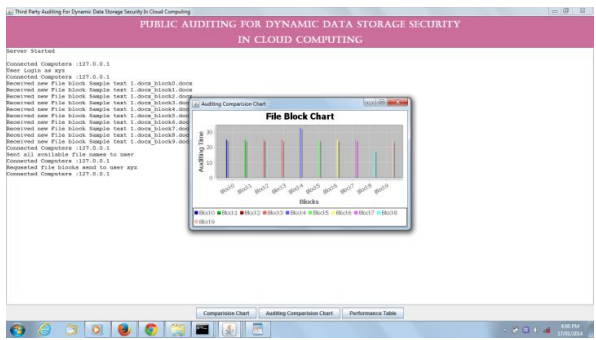

Fig.8 shows metadata into the block with auditin time.

## VII. CONCLUSION

Our system ensures remote data integrity with support of both public audit ability and dynamic data operations.

Considering TPA may concurrently handle multiple audit sessions from different user's for their outsourced data files as well as support for data dynamic operation such as block modification, insertion, deletion, verification and taking log history of client performed operation. also the acknowledgement is getting to cloud space user from the server by the TPA.